\documentclass[prb,floatfix,twocolumn,showpacs,amsmath,amssymb]{revtex4-2}
\usepackage{graphicx}
\usepackage{bm}
\usepackage{color}
\usepackage{txfonts}

\begin{document}

\title{Crossover in Electronic Specific Heat near Narrow-Sense Type-III Dirac Cones}




\author{Keita Kishigi}
\affiliation{Faculty of Education, Kumamoto University, Kurokami 2-40-1, Kumamoto, 860-8555, Japan}

\author{Yasumasa Hasegawa}
\affiliation{Department of Material Science, Graduate School of Material Science, University of Hyogo, Hyogo, 678-1297, Japan}

\begin{abstract}

Two-dimensional massless Dirac fermions exhibit Dirac cones, which are classified into three types: type-I, type-II, and type-III. In both type-I and type-II cones, the energy dispersion is linear in all momentum directions. Type-I cones are characterized by a non-overtilted structure, where the Dirac point serves as a local minimum (maximum) for the upper (lower) band. In contrast, type-II cones exhibit overtilted dispersions, leading to the coexistence of electron and hole pockets. At the critical tilt, the linear energy dispersion vanishes in one momentum direction, corresponding to a type-III Dirac cone. We further define a special case, termed the ``narrow-sense'' type-III cone, where not only the linear term but also quadratic and higher-order terms vanish, resulting in a completely flat dispersion along one direction. In this work, we numerically investigate the temperature ($T$) -dependence of the electronic specific heat ($C$), as the Dirac cone is continuously tilted from type-I to narrow-sense type-III. A model with particle-hole symmetry is employed to ensure that the chemical potential ($\mu$) remains temperature independent. Our results reveal a notable crossover in $C$ near narrow-sense type-III, where $C$ changes from $C \propto T^{2}$ below the crossover temperature ($T_{\rm co}$) to $C \propto T^{\frac{1}{2}}$ above $T_{\rm co}$. This crossover is attributed to the energy-dependent structure of the density of states. The present findings suggest a feasible approach for experimentally probing the degree of Dirac cone tilting near the narrow-sense type-III limit.

\end{abstract}


\date{\today}


\maketitle

\section{Introduction}\label{intro}

Two-dimensional (2D) massless Dirac fermion systems are realized in materials such as graphene\cite{novo2005}, $\alpha$-(BEDT-TTF)$_2$I$_3$\cite{Kajita2014,Hirata2011,Osada2008,Konoike2012}, and $\alpha$-(BETS)$_2$I$_3$\cite{Tajima2021,kitou2021}, among others, where Dirac points and Dirac cones are present. These systems exhibit a variety of intriguing properties, particularly when two Dirac points 
merge\cite{Tarruell2012,Bellec,Mil,real,Shao}.

One notable phenomenon is a topological phase transition to a normal insulating state that occurs upon the merging of two Dirac points. This has been theoretically predicted in systems such as honeycomb lattices\cite{Hasegawa2006,Dietl2008}, VO$_2$/TiO$_2$ nanostructures\cite{Bane}, and  $\alpha$-(BEDT-TTF)$_2$I$_3$\cite{Suzumura2013}. This merging point is termed the semi-Dirac point\cite{Bane}. 
Experimental realizations of this transition and a semi-Dirac point have been reported in artificial systems including ultracold atoms in optical lattices~\cite{Tarruell2012}, photonic resonator lattices~\cite{Bellec}, and polariton systems in semiconductor micropillar lattices~\cite{Mil,real}. More recently, evidence of a semi-Dirac point has been reported in real materials such as ZrSiS~\cite{Shao}. 


Near the merging of two Dirac points, a crossover behavior in the temperature ($T$)-dependence of the electronic specific heat $(C)$ has been identified through numerical calculations~\cite{KTH2023}. Specifically, $C \propto T^2$ at low temperatures and $C \propto T^{3/2}$ at higher temperatures, separated by a crossover temperature ($T_{\rm co}$). This crossover is caused by the enegry ($\varepsilon$)-dependence of the density of states (DOS) [$D(\varepsilon)$], where $D(\varepsilon) \propto |\varepsilon-\varepsilon_{\rm F}|$ near the Fermi energy ($\varepsilon_{\rm F}$)  (as expected for 2D Dirac systems) and $D(\varepsilon) \propto |\varepsilon-\varepsilon_{\rm F}|^{{1/2}}$ far from $\varepsilon_{\rm F}$ (as expected for 2D semi-Dirac systems). As two Dirac points approach each other, $T_{\rm co}$ shifts to lower values and eventually vanishes upon merging. This crossover thus serves as an important indicator of a topological phase transition in 2D Dirac systems\cite{KTH2023}.

The experimental observation of the  electronic specific heat, $C$, has been achieved\cite{Konoike2012} only in bulk materials such as $\alpha$-(BEDT-TTF)$_2$I$_3$, where $C\propto T^{1.8}$ has been measured and is close to expected $C\propto T^{2}$. However, the crossover in $C$ associated with the topological phase transition has not yet been observed. It may be detectable under uniaxial strain.

\begin{figure}[bt]
\begin{flushleft} \hspace{0.5cm}(a) \end{flushleft}\vspace{-0.2cm}\hspace{0.2cm}
\includegraphics[width=0.43\textwidth]{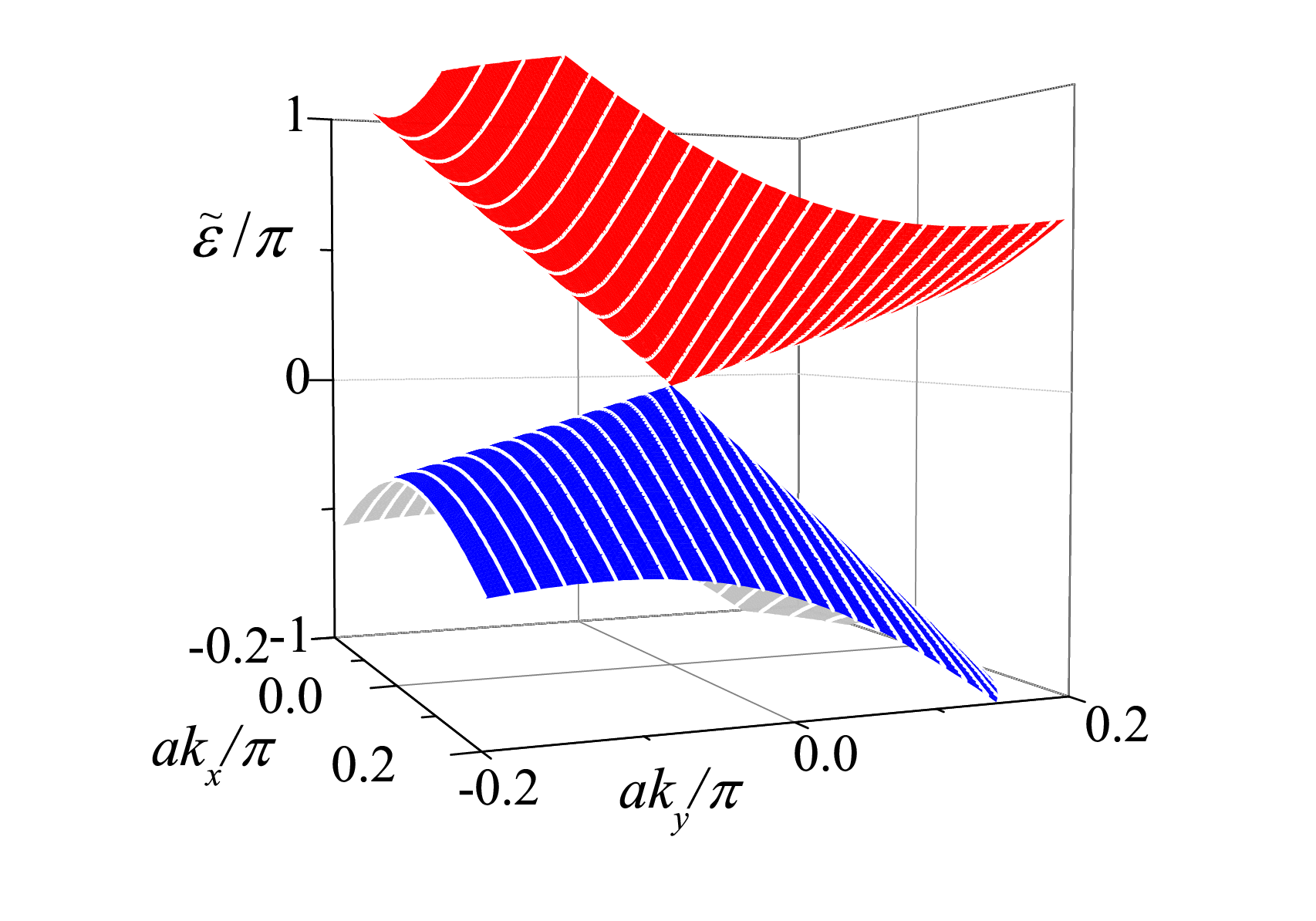}\vspace{-0.2cm}
\begin{flushleft} \hspace{0.5cm}(b) \end{flushleft}\vspace{-0.2cm}\hspace{0.2cm}
\includegraphics[width=0.43\textwidth]{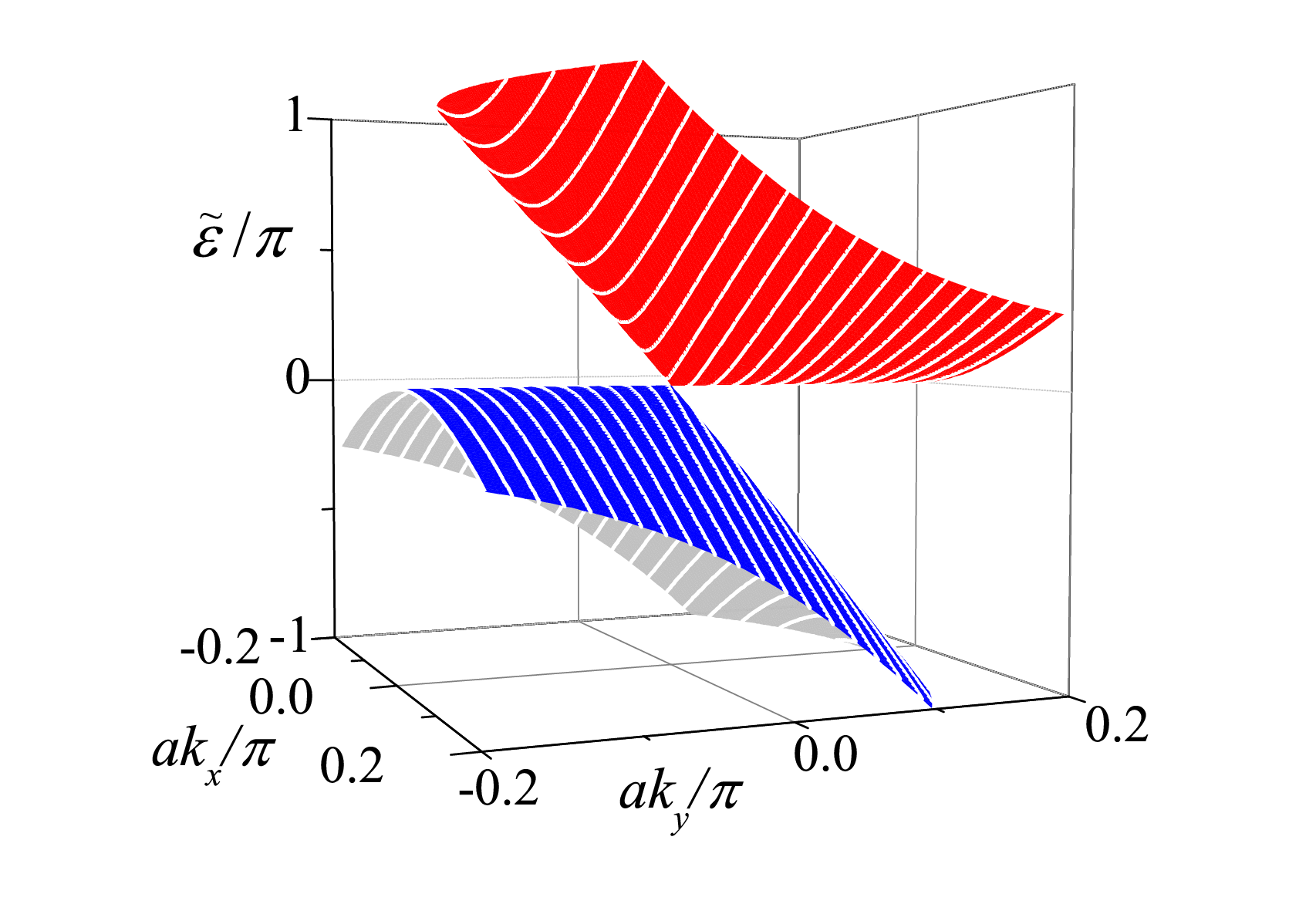}\vspace{-0.2cm}
\begin{flushleft} \hspace{0.5cm}(c) \end{flushleft}\vspace{-0.2cm}\hspace{0.2cm}
\includegraphics[width=0.43\textwidth]{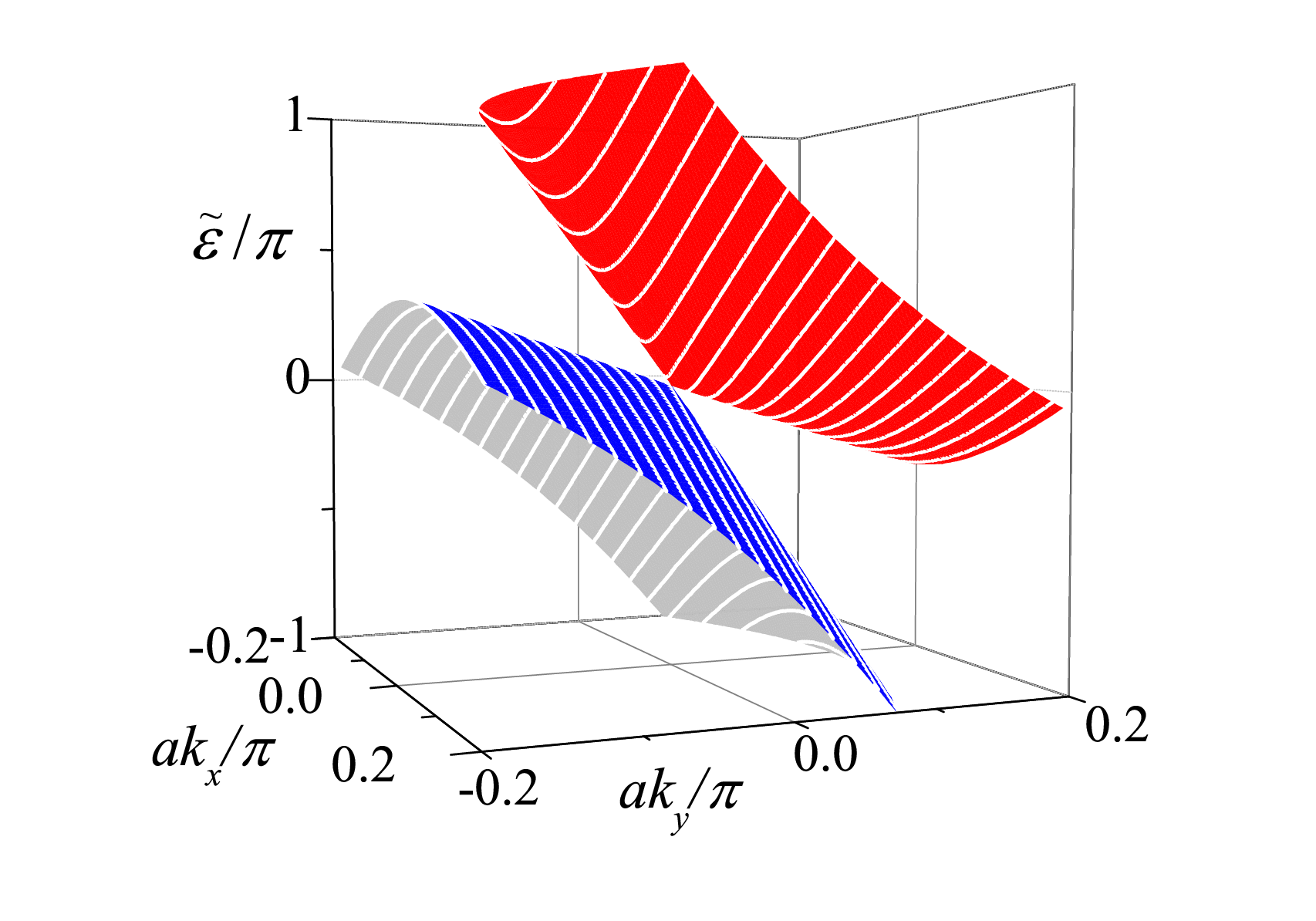}\vspace{-0.2cm}
\caption{(Color online) 
The energy bands of 2D type-I (a), narrow-sense type-III (b), and type-II (c) Dirac described by Eq. (\ref{model}), where  
we take $\varepsilon_{\rm F}=0$ and ${\tilde \alpha}=-0.5$ for (a), ${\tilde \alpha}=-1$ for (b), and ${\tilde \alpha}=-1.5$ for (c).  The Dirac point ($\mathbf{k}_{\rm D}$) is $\mathbf{k}_{\rm D}=(0,0)$ and the Dirac cone is tilted to the 
$k_y$-axis.
}
\label{fig00}
\end{figure}

\begin{figure}[bt]
\vspace{0.2cm}
\includegraphics[width=0.48\textwidth]{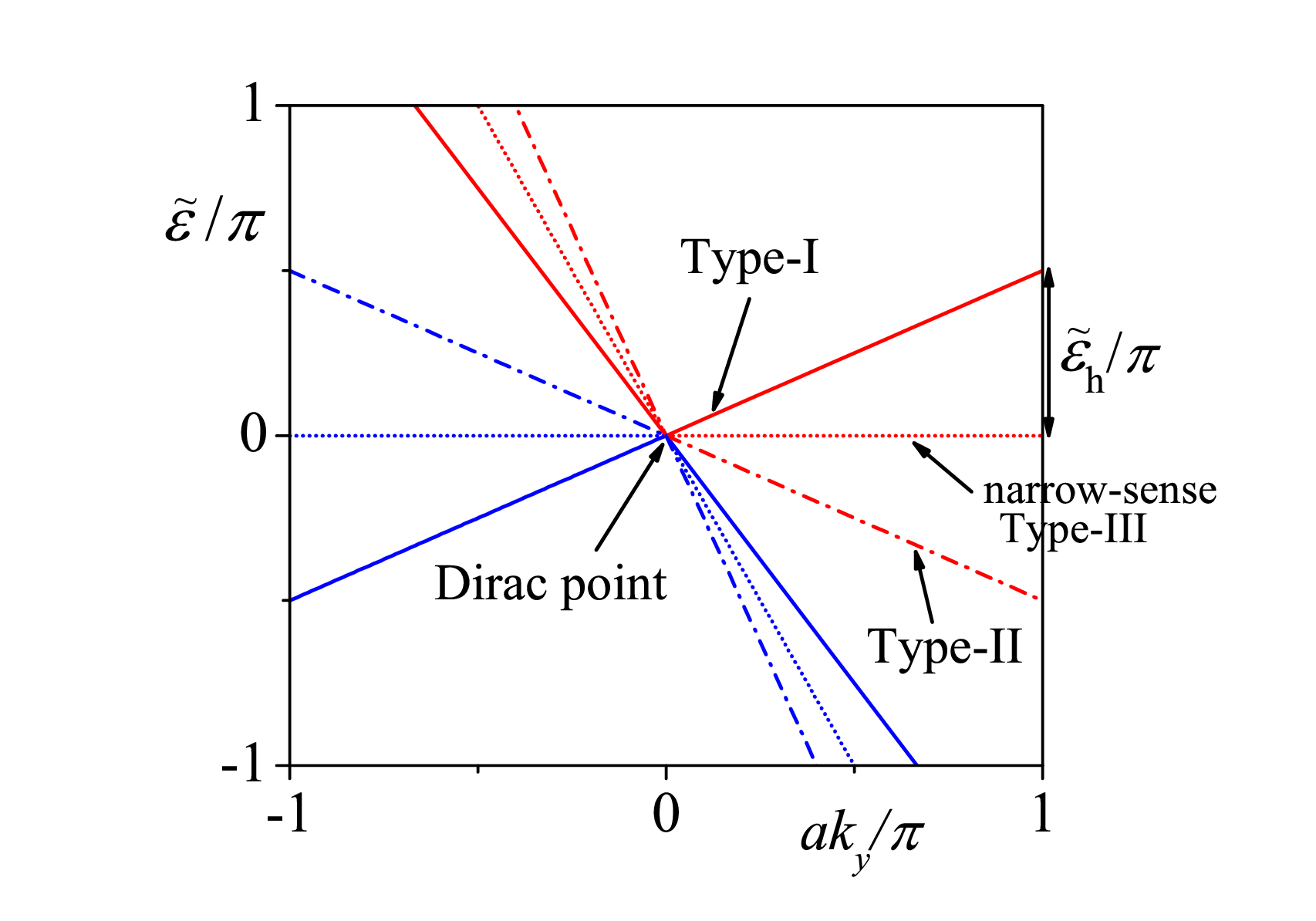}\vspace{-0.0cm}
\caption{(Color online) 
The energy dispersions of type-I (solid lines), narrow-sense type-III (dotted lines), and type-II (dashed-dotted lines) shown in Figs.~\ref{fig00}(a), (b), and (c), respectively, are plotted from the viewpoint along the $k_x$-axis. The dimensionless height of the Dirac cone (${\tilde \varepsilon_{\rm h}}$) is defined by the
 height from $k_y=0$ to $\pi/a$ (or $-\pi/a$) in the coordinate system used in this figure. We obtain ${\tilde \varepsilon_{\rm h}}=\pi({\tilde \alpha}\pm1)$, where $+$ and $-$ signs correspond to ${\tilde \alpha} < 0$ and ${\tilde \alpha} > 0$, respectively. 
For type-I with ${\tilde \alpha}=-0.5$, ${\tilde \varepsilon_{\rm h}} =0.5$. For ${\tilde \alpha}=1$ or $-1$, the Dirac cone becomes narrow-sense type-III. 
}
\label{fig0}
\end{figure}



The concept of a tilted Dirac cone has been extensively studied in 2D massless Dirac systems, particularly in $\alpha$-(BEDT-TTF)$_2$I$_3$, both theoretically~\cite{katayama2006,kino} and experimentally~\cite{hirata2016}. Tilted cones are typically classified into type-I, -II, and -III. Type-I cones are tilted but not overtilted, type-II are overtilted, and type-III are critically tilted where the linear dispersion vanishes along one-direction. Type-III Dirac cones have been proposed in systems including Zn$_2$In$_2$\cite{huang2018}, laser-irradiated  black phosphorus\cite{Liu}, and Ni$_3$In$_2$$X_2$ ($X$ = S, Se)\cite{sims}, based on first-principles calculations.

Recently, type-III has drawn attention due to its analogy to black holes in condensed matter systems\cite{vol1,vol2,vol3}. Theoretical studies have revealed that type-III systems exhibit unique transport properties distinct from those of types-I and -II~\cite{mizoguchiPRB}. 

\begin{figure}[bt]
\vspace{0.2cm}
\includegraphics[width=0.48\textwidth]{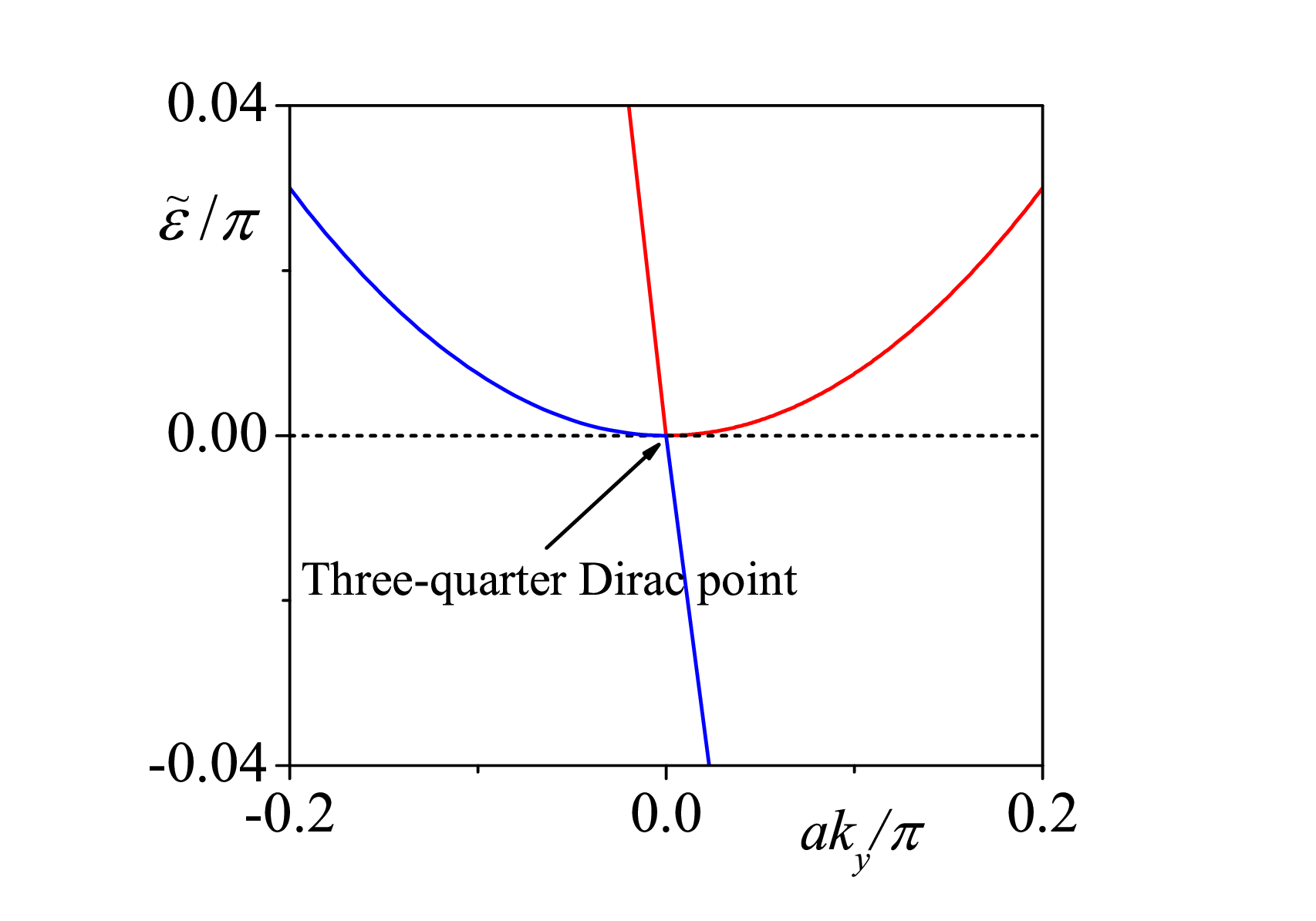}\vspace{-0.0cm}
\caption{(Color online) 
The energy dispersion of three-quarter Dirac described by Eq. (\ref{model_tq}) with ${\tilde \alpha} = -1$, $\varepsilon_{\rm F}=0$, $\alpha^{\prime}_2=\hbar v_{\rm F}/2$, and $\alpha^{\prime\prime}_2=\hbar v_{\rm F}/4$ is viewed along the $k_x$-axis.
}
\label{fig3}
\end{figure}


In this study, we focus on the electronic specific heat in the vicinity of ``narrow-sense type-III'' Dirac cones, where the linear and higher-order terms vanishes.  We adopt a minimal model given by
\begin{equation}
 \varepsilon = \varepsilon_{\rm F} + \alpha k_y \pm \hbar v_{\rm F} \sqrt{k_x^2 + k_y^2}, \label{model}
 \end{equation} 
where $h$ is Planck's constant, $\hbar = h/(2\pi)$, $v_{\rm F}$ is the Fermi velocity, and the parameter, $\alpha$,  characterizes the tilting of the Dirac cone along the $k_y$ direction. In this study, the wave vector space is restricted to the range, $-\pi/a < k_x \leq \pi/a$ and $-\pi/a < k_y \leq \pi/a$. We define a dimensionless tilting parameter as
$\tilde{\alpha} = \alpha/(\hbar v_{\rm F})$, and the system is classified as type-I for $|\tilde{\alpha}| < 1$, narrow-sense type-III for $|\tilde{\alpha}| = 1$, and type-II for $|\tilde{\alpha}| > 1$. Figures~\ref{fig00} and \ref{fig0} show the energy dispersion for representative values of $\tilde{\alpha}$ in each regime. In the main text, we use the dimensionless energy, ${\tilde \varepsilon} = a\varepsilon/(\hbar v_{\rm F})$ and the  dimensionless Fermi energy, ${\tilde \varepsilon} _{\rm F}= a\varepsilon_{\rm F}/(\hbar v_{\rm F})$.

In the narrow-sense type-III case, the dispersion becomes flat along the $k_y$ direction at $k_x = 0$. In this work, the Fermi energy $\varepsilon_{\rm F}$ is set to the Dirac point in the type-I case and to the flat band segment in the narrow-sense type-III case. Accordingly, the electron filling is fixed at half-filling, $\nu = 1/2$.



To incorporate higher-order effects near the critical tilting, we also consider a modified model:
\begin{equation}
\varepsilon= \varepsilon_{\rm F}+\alpha k_y+\alpha_2^{\prime} k_y^2\pm\hbar v_{\rm F}\sqrt{k_x^2+(\alpha k_y-\alpha^{\prime\prime}_2k_y^2)^2}, \label{model_tq}
\end{equation}
where $\alpha^{\prime}_2$ and/or $\alpha^{\prime\prime}_2$ are nonzero for $|{\tilde \alpha}|=1$. This model exhibits  a three-quarter-Dirac cone\cite{KH2017}, characterized by quadratic dispersion in one direction and linear dispersion in the other three directions. Such a feature appears in the tight-binding model for $\alpha$-(BEDT-TTF)$_2$I$_3$ under a uniaxial pressure of 2.3 kbar\cite{KH2017}. A three-quarter-Dirac cone can be regarded as a more general form of the type-III Dirac cone, because the energy dispersion exhibits locally flat behavior at the three-quarter-Dirac point along the $k_y$-axis, as shown in  Fig. \ref{fig3}. 

The electronic structures of narrow-sense type-III and three-quarter Dirac cones exhibit fundamentally different magnetic and thermodynamic properties. For instance, the Landau quantization of the upper band in the three-quarter Dirac system follows 
$E_n \propto B^{4/5}$\cite{KH2017}, whereas in narrow-sense type-III it scales as $B^2$. In this work, we restrict our analysis to the narrow-sense type-III case based on Eq. (\ref{model}), to clarify the specific heat behavior near the critical tilting.

Importantly, while Eq.~(\ref{model}) applies for type-II ($|\tilde{\alpha}| > 1$), in that regime electron and hole pockets typically emerge, making the linear approximation invalid. We therefore limit our analysis of $C$ to the vicinity of the type-I/type-III boundary.

In 2D systems for type-I and narrow-sense type-III, the DOS is given by (see Appendix \ref{appenA})
\begin{eqnarray}
D(\varepsilon)_{}\propto&  
 \left\{ \begin{array}{llll}
|{\tilde \varepsilon}-{\tilde \varepsilon}_{\rm F}|  & \mbox{for type-I Dirac},\\
\ |{\tilde \varepsilon}-{\tilde \varepsilon}_{\rm F}|^{-1/2} & \mbox{for narrow-sense type-III Dirac}.\\
\end{array} \right. 
\label{dos_special}
\end{eqnarray}
The DOS  diverges at ${\tilde \varepsilon}={\tilde \varepsilon}_{\rm F}$ in narrow-sense type-III, because the electronic states are concentrated  in the flat dispersion at $k_x=0$.



\begin{figure}[bt]
\vspace{-0.0cm}
\begin{flushleft} \hspace{0.5cm}(a) \end{flushleft}\vspace{-0.2cm}
\includegraphics[width=0.50\textwidth]{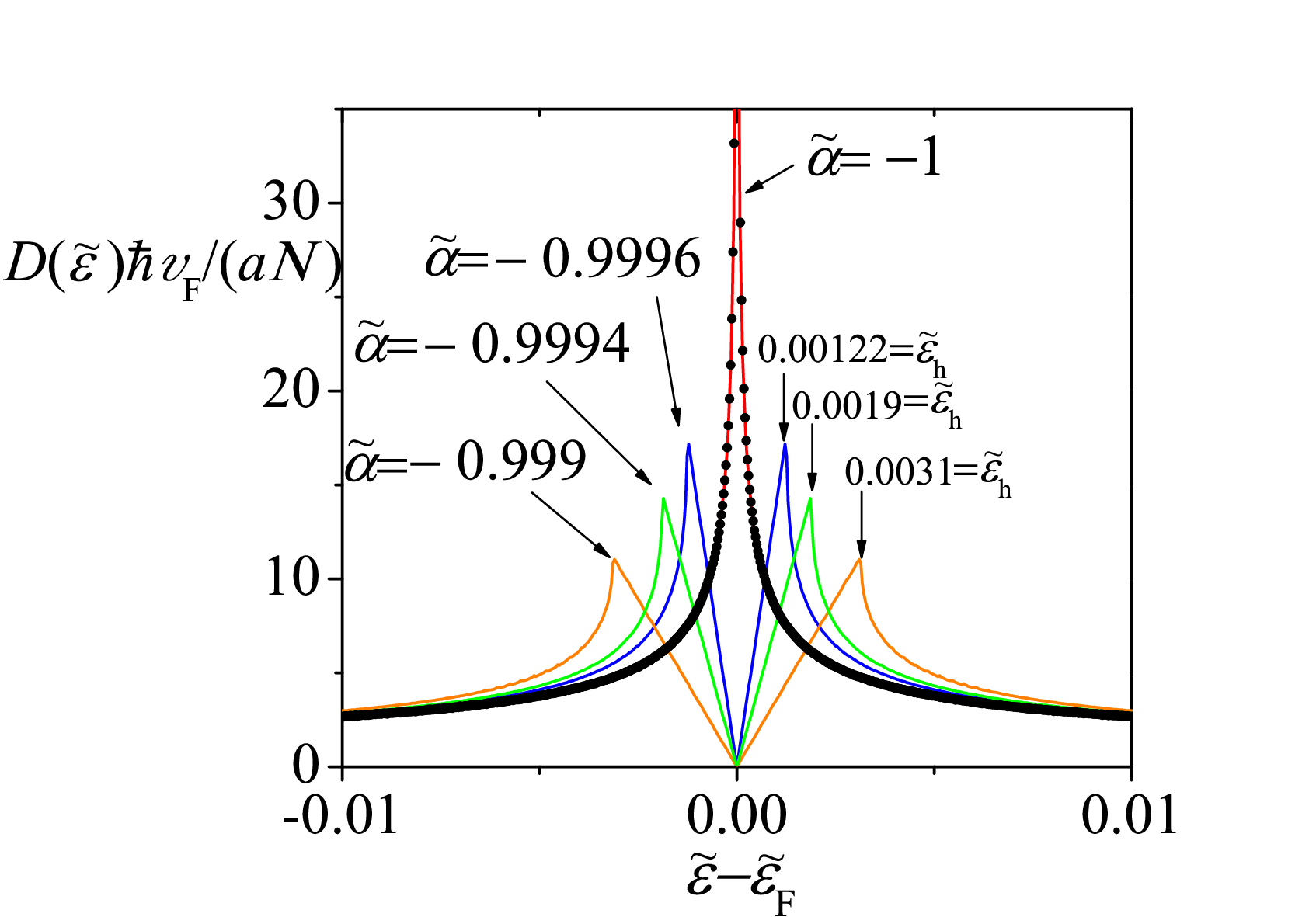}\vspace{-0.0cm}
\begin{flushleft} \hspace{0.5cm}(b) \end{flushleft}\vspace{-0.3cm}
\includegraphics[width=0.50\textwidth]{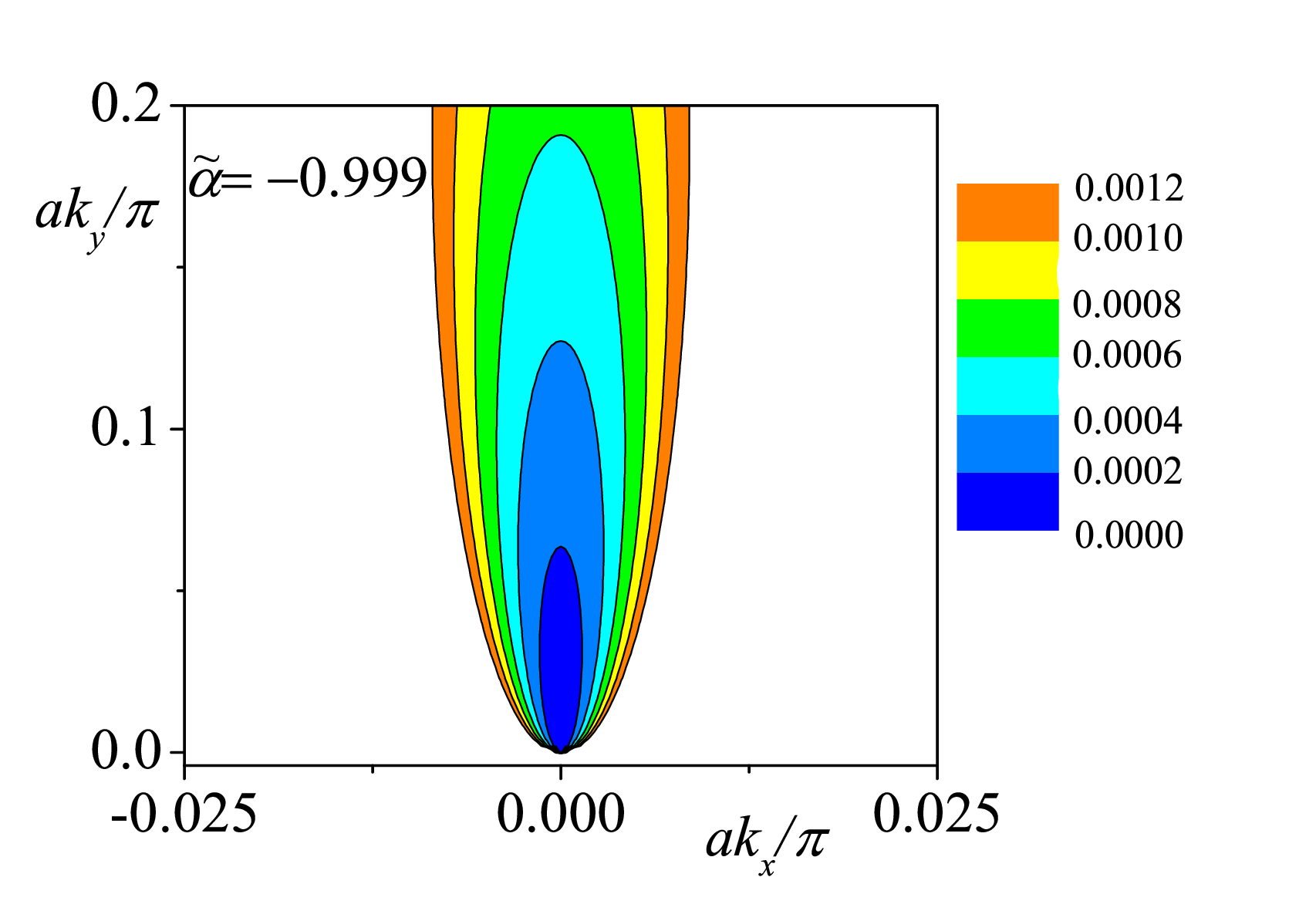}\vspace{-0.0cm}
\caption{(Color online) 
 (a) The DOS for Dirac systems evolving from type-I to narrow-sense type-III regimes. The blue, green, and orange curves correspond to numerical calculations using Eq. (\ref{model}) with ${\tilde \alpha} = -0.9996$, $-0.9994$, and $-0.999$, respectively. Black circles represent the numerical result for ${\tilde \alpha} = -1$, obtained within the first Brillouin zone defined by $-\pi/a < k_x \leq \pi/a$ and $-\pi/a < k_y \leq \pi/a$. The red solid line denotes the analytical expression given by Eq. (\ref{d_0_3}), which is in excellent agreement with the numerical data for ${\tilde \alpha} = -1$.
(b) Contour plot of the upper energy band described by Eq. (\ref{model}) for ${\tilde \alpha} = -0.999$ at $\varepsilon_{\rm F} = 0$. Near $\varepsilon \simeq \varepsilon_{\rm F}$, the electronic states are localized around two Dirac points. For higher energies (e.g., $\varepsilon \gtrsim 0.0008$), the states extend along a line-shaped region centered around $k_x \simeq 0$ and $0 < k_y \leq \pi/a$. A similar behavior is observed in the lower band for $\varepsilon \ll \varepsilon_{\rm F}$, where states are distributed along a line near $k_x \simeq 0$ and $-\pi/a < k_y < 0$. These features underlie the characteristic  energy dependence of the DOS for ${\tilde \alpha} = -0.999$ shown in panel (a).
}
\label{fig_dos}
\end{figure}

When the Dirac cone is tilted from the type-I toward the narrow-sense type-III regime, the behavior of the DOS, $D(\varepsilon)$, is shown in Fig.~\ref{fig_dos}(a). For $\tilde{\alpha} =-0.9996, -0.9994$, and $-0.999$, $D(\tilde{\varepsilon})$ exhibits a linear dependence near the Fermi energy, $D(\tilde{\varepsilon}) \propto |\tilde{\varepsilon} -\tilde{\varepsilon}_{\rm F}|$, and $D(\tilde{\varepsilon})$ has the maximum at 
$|\tilde{\varepsilon} - \tilde{\varepsilon}_{\rm F}| = \tilde{\varepsilon}_h$, where $\tilde{\varepsilon}_h$ is maximum energy from the Dirac energy (i.e., $\tilde{\varepsilon}_{\rm F}$) in the tilted direction  (see the caption of Fig. \ref{fig0}). 
The DOS behaves approximately as $D(\tilde{\varepsilon})$$\propto |\tilde{\varepsilon} - \tilde{\varepsilon}_{\rm F}|^{-1/2}$ when  
$|\tilde{\varepsilon} - \tilde{\varepsilon}_{\rm F} | > \tilde{\varepsilon}_h$. This behavior originates from our restriction of  the wave vector space ($-\pi/a < k_x \leq \pi/a$ and $-\pi/a < k_y \leq \pi/a$). Within this simple model, saddle points in the energy dispersion and the associated logarithmic divergence in the DOS are absent. Nevertheless, the $T$-dependence of $C$ remains essentially unaffected.

Accordingly, a thermal crossover in $C$ is observed near the narrow-sense type-III regime, from $C \propto T^{2}$ at low temperatures ($T < T_{\rm co}$) to $C \propto T^{1/2}$ at higher temperatures ($T > T_{\rm co}$). The crossover temperature, $T_{\rm co}$, is found to be of the same order as, but smaller than, $\tilde{\varepsilon}_{\rm h}$. This is because the function $-\partial f / \partial \varepsilon$ has a characteristic width on the order of $T$, effectively broadening the contribution of the DOS over an energy range $\sim T$. Consequently, even if the DOS exhibits a sharp feature at a certain energy, its effect on the specific heat manifests at temperatures somewhat below that energy scale.


\section{Calculation of Electronic Specific Heat}
\label{esh}

While Coulomb interactions\cite{Vaf,She,Kot} have been theoretically accounted for in graphene, such as demonstrating logarithmic corrections to thermodynamic properties[30], we assume that the influence of Coulomb interactions is negligible
compared to that of the tilting of Dirac cones.

The internal energy per site, $U$, at temperature, $T$, is given by
\begin{equation} U= \frac{1}{N} \sum_{i = \pm} \sum_{\mathbf{k}} \varepsilon_i(\mathbf{k}) f[\varepsilon_i(\mathbf{k})], \label{U_1} 
\end{equation}
where 
$\varepsilon_i({\bf k})$ denotes the energy eigenvalue of band $i$, and $f[\varepsilon_i({\bf k})]$ is the Fermi-Dirac distribution function. Here, 
$N=2N_k$ with $N_k$ representing the number of $\mathbf{k}$ points sampled within the first Brillouin zone.

The electronic specific heat at constant volume, $C$, is obtained by taking the temperature derivative of the internal energy: 
\begin{eqnarray}
C=\frac{dU}{dT}. \label{Ce}
\end{eqnarray}
In this study, we ignore the spin degree of freedom, which does not affect the power-law behavior of  $C$ with respect to $T$. 
 Including spin would simply introduce a factor of 2 on the right-hand side of Eq. (\ref{Ce}).

\section{Analytic Calculations in Special Cases with particle-hole symmetry}

We consider the following generalized DOS with particle-hole symmetry:
\begin{align}
D(\varepsilon)=ND_{0}|\varepsilon-\varepsilon_{\rm F}|^{\beta}, \label{D1}
\end{align}
where $D_{0}$ and $\beta$ are constant. For Eq. (\ref{D1}), we derive an analytical expression for $C$ in Appendix \ref{appenB}, which is presented in Eq.~(\ref{CT}). 
This leads to characteristic power-law $T$-dependences of $C$ as follows:
\begin{eqnarray}
C_{}\propto&  
 \left\{ \begin{array}{llll}
{T}^{2} & \mbox{for type-I Dirac},\\
{T}^{1/2} & \mbox{for narrow-sense type-III Dirac}.\\
\end{array} \right. 
\end{eqnarray}




\begin{figure}[bt]
\begin{flushleft} 
\vspace{0.2cm}
\hspace{0.2cm}
(a) \end{flushleft}
\includegraphics[width=0.55\textwidth]{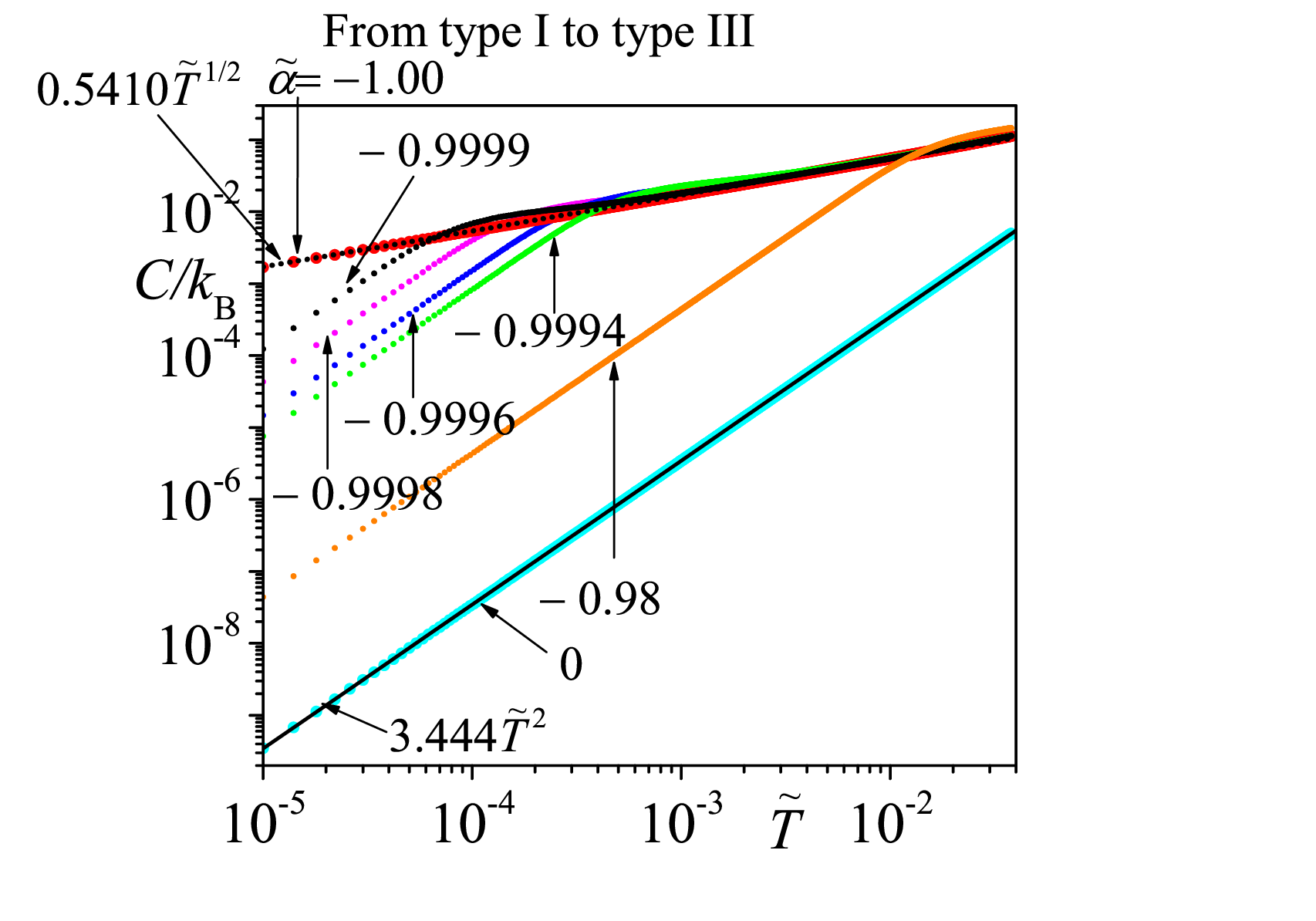}\vspace{-0.2cm}
\begin{flushleft} \hspace{0.2cm}(b) \end{flushleft}\vspace{-0.3cm}\hspace{0.2cm}
\includegraphics[width=0.55\textwidth]{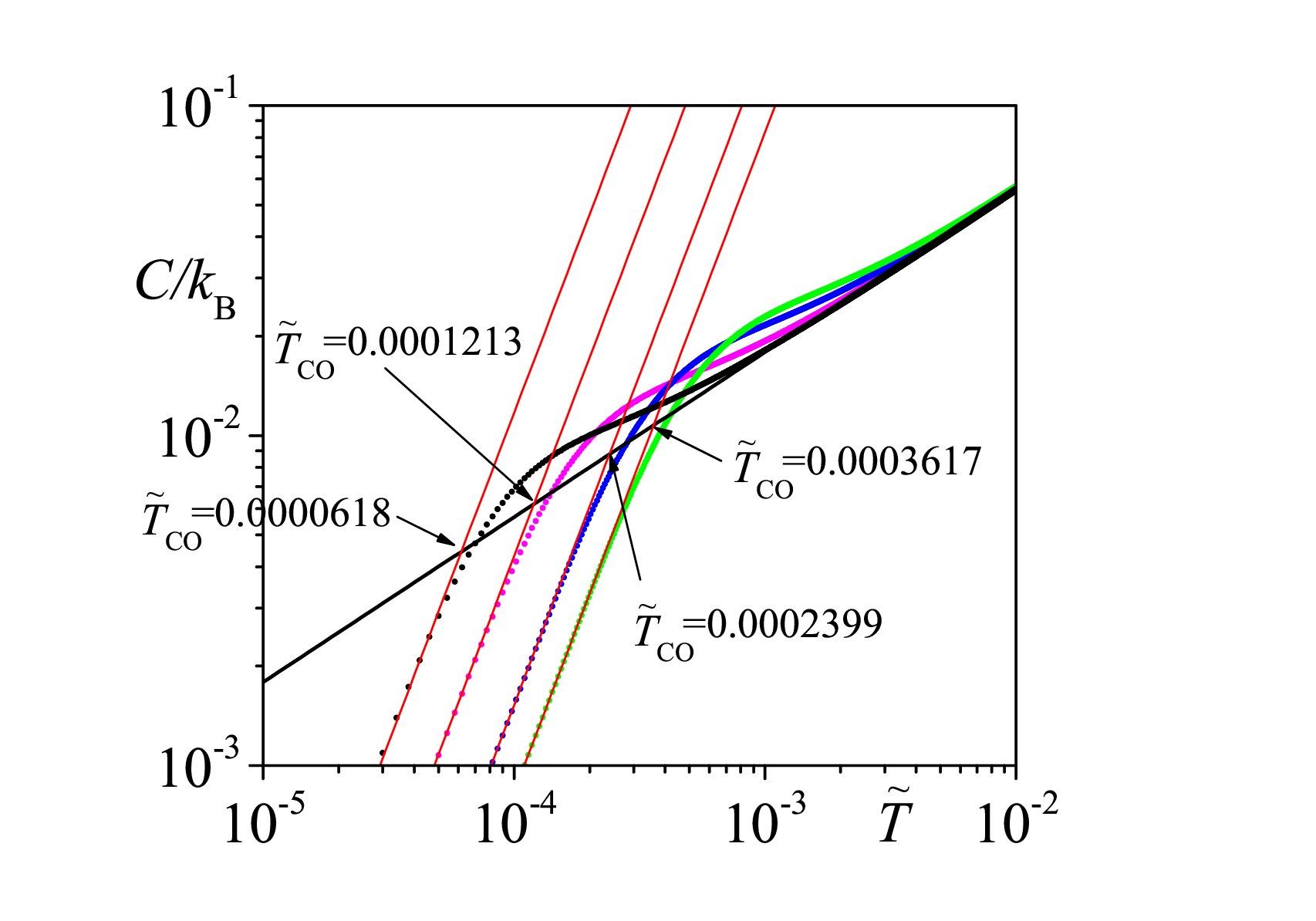}\vspace{-0.2cm}
\caption{(Color online) (a) 
Electronic specific heats, calculated numerically  from Eq. (\ref{model}), are shown as red circles and other colored circles 
(black, pink, blue, green, orange, sky blue) corresponding to values of 
${\tilde \alpha}=-1$, $-0.9999, -0.9998, -0.9996, -0.9994, -0.98$, and $0$.  
A dotted black line and a black solid line represent $0.5410{\tilde T}^{1/2}$ from Eq. (\ref{type3}) and  3.444${\tilde T}^{2}$ from Eq. (\ref{C_D}), respectively. (b) The estimated crossover temperatures, ${{\tilde T}}_{\rm co}$, are 0.0000618, 0.0001213, 0.0002399, and 0.0003617 for ${\tilde \alpha}$=$-0.9999, -0.9998, -0.9996$, and $-0.9994$.
}
\label{fig_12}
\end{figure}


\begin{figure}[bt]
\vspace{0.3cm}
\includegraphics[width=0.55\textwidth]{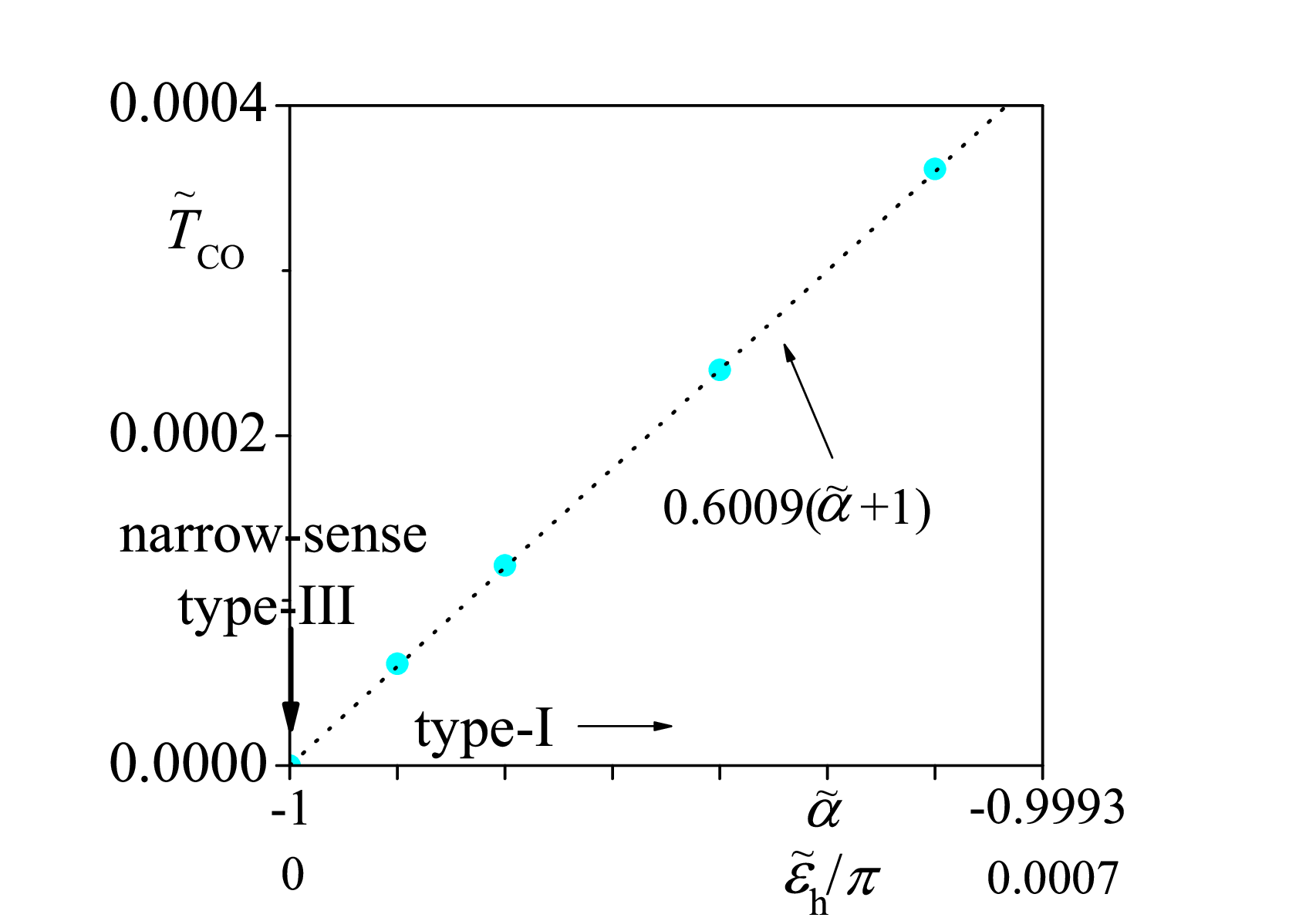}\vspace{0.2cm}
\caption{(Color online) 
Crossover temperatures as a function of ${\tilde \alpha}$ (or $\theta$) from type-I to narrow-sense type-III (sky blue circles). The sky blue circles  are fitted well as ${{\tilde T}}_{\rm co}=0.6009({\tilde \alpha} +1)$ or ${{\tilde T}}_{\rm co}=0.1913{\tilde \varepsilon_{\rm h}}$.
}
\label{fig_7}
\end{figure}

\section{Numerical Calculations for Dirac cone tilting from type I to type III}
\label{ES_2D}



Since the Dirac points appear as a pair at the time-reversal-invariant-momenta in general, we multiply the numerically obtained $C$ by a factor of 2. The momentum summation is performed over the first Brillouin zone of a square lattice, defined by $-\pi/a\leq k_x<\pi/a$ and $-\pi/a\leq k_y<\pi/a$. The power-law behavior of $C$ as a function of $T$ remains unchanged even if any Brillouin zone is chosen. 


Throughout this work, we employ a dimensionless temperature defined as
\begin{equation}
\tilde{T}=\frac{ak_{\rm B}T}{\hbar v_{\rm F}}, 
\end{equation}
where $k_{\rm B}$ is the Boltzmann constant. For graphene, it has been known\cite{wallace} that $\hbar v_{\rm F}=\sqrt{3}at/2$, 
where $t$ is nearest transfer integrals. 
Based on first-principles calculations\cite{Reich}, $t$ is estimated to be approximately 2.97 eV, implying that $\tilde{T}=0.0001$ corresponds to approximately 2.98 K. 

Figures \ref{fig_12} displays the numerically calculated $C$. For type-I (${\tilde \alpha}=0$), $C$ (sky blue circles) is well fitted by the analytical expression in Eq. (\ref{C_D}) (black solid line). Similarly, for narrow-sense type-III (${\tilde \alpha}=-1$), $C$ (red circles) is accurately fitted by Eq. (\ref{type3}) (dotted black line).

Near the narrow-sense type-III regime ($-0.9999\leq {\tilde \alpha}\leq -0.98$), a crossover behavior is observed in the $T$-dependence of $C$. 
At low temperatures, $C\propto {\tilde T^2}$, as expected for type-I, while at higher temperatures, it transitions to $C\propto {\tilde T}^{1/2}$, characteristic of the narrow-sense type-III.  This crossover arises from the change in $\varepsilon$-dependence of $D(\varepsilon)$ (see Fig. \ref{fig_dos}). The low temperature behavior of $C$ is primarily governed by  $D(\varepsilon)$ near $\varepsilon_{\rm F}$, 
while the high temperature behavior of $C$ reflects contributions from states farther away from $\varepsilon_{\rm F}$.


At high temperatures, the electronic specific heat $C$ for $\tilde{\alpha} = -0.9999$, $-0.9998$, $-0.9996$, and $-0.9994$ converges to that for $\tilde{\alpha} = -1$, which corresponds to the narrow-sense type-III Dirac cone, as shown in Fig.\ref{fig_12}(a). We define the crossover temperature $\tilde{T}_{\rm co}$ as the temperature at which the red lines---obtained by extrapolating $C$ from low-temperature data ($\tilde{T} = 0.00001$ to $0.00005$) for $\tilde{\alpha} = -0.9999$, $-0.9998$, $-0.9996$, and $-0.9994$---intersect with the black line, which is extrapolated from high-temperature data ($\tilde{T} = 0.01$ to $0.03$) for $\tilde{\alpha} = -0.9999$, as shown in Fig. \ref{fig_12}(b). 

As shown in Fig. \ref{fig_7},  ${{\tilde T}}_{\rm co}$ increases linearly with ${\tilde \alpha}$, or equivalently indicating that ${{\tilde T}}_{\rm co}=0.1913{\tilde \varepsilon_{\rm h}}$. One can estimate the characteristic energy scale at which the DOS changes from the observed value of 
${\tilde T}_{\rm co}$.

\section{Conclusions}

When the Dirac cone is tilted from type-I to narrow-sense type-III in 2D Dirac fermions, a notable crossover emerges in the ${\tilde T}$-dependence of the electronic specific heat, $C$. 
Assuming particle-hole symmetry is preserved (or its breaking does not significantly influence the ${\tilde T}$-dependence of the chemical potential, $\mu$), the specific heat exhibits a transition from  $C\propto {\tilde T}^{2}$ at low temperatures (${\tilde T}<{\tilde T}_{\rm co}$) to $C\propto {\tilde T}^{1/2}$ at higher temperatures (${\tilde T}>{\tilde T}_{\rm co}$). This crossover arises from the modification of the low-energy DOS due to the tilting of the Dirac cone. Importantly, it is not the result of finely tuned model parameters but represents a generic and robust feature of tilted Dirac systems approaching the narrow-sense type-III limit. The crossover temperature, ${\tilde T}_{\rm co}$, 
is proportional to ${\tilde \varepsilon_{\rm h}}$ and approaches zero in the narrow-sense type-III limit. This behavior offers a practical way to experimentally probe Dirac cone tilting near narrow-sense type-III systems when the Fermi energy lies at the Dirac point and the ${\tilde T}$-dependence of $\mu$ is negligibly small.

In our model, particle-hole symmetry is preserved. In contrast, in certain Dirac systems such as the tight-binding model on the honeycomb lattice including next-nearest neighbor hoppings\cite{kishigi2011}, three-quarter-Dirac, and $\alpha$-(BEDT-TTF)$_2$I$_3$\cite{KTH2023}, this symmetry is  significantly broken. In such cases, the 
${\tilde T}$-dependence of  $\mu$ pronounced, even when the Dirac point lies at the Fermi energy. It is therefore essential to properly account for this effect when evaluating the specific heat, $C$. 


When the filling is slightly away from half-filling in our model, one still expects crossovers in the specific heat to behaviors such as ${\tilde T}^2$ or ${\tilde T}^{1/2}$ if the ${\tilde T}$-dependence of $\mu$ can be neglected, However, when the deviation from half-filling becomes large 
($|{\tilde \varepsilon}-{\tilde \varepsilon}_{\rm F}|\gtrsim 0.01$), the DOS near $\mu$ becomes nearly constant, which may give rise to a different crossover to $C \propto {\tilde T}$ behavior. These considerations suggest that tuning the filling could lead to rich and intriguing phenomena. Nevertheless, away from half-filling, the ${\tilde T}$-dependence of $\mu$ becomes significant, and its effect must be incorporated in evaluating the specific heat. A detailed investigation of these effects is left for future work.

Recent first-principles density functional theory calculations have revealed that $\alpha$-(BETS)$_2$I$_3$ hosts a nearly flat Dirac cone with the Fermi energy lying close to the Dirac point\cite{kitou2021}. This characteristic suggests that $\alpha$-(BETS)$_2$I$_3$ is a viable candidate for experimentally observing the specific heat crossover predicted in our study. In particular, when the Dirac cone in this material closely approximates the narrow-sense type-III cone and the ${\tilde T}$-dependence of $\mu$ remains weak at low temperatures, our theoretical results are directly applicable. These conditions provide an opportunity to test the universal crossover behavior we propose, using an experimentally accessible system.

To fully understand the thermal and magnetic properties of type-III Dirac systems, a clear distinction between narrow-sense type-III and three-quarter-Dirac dispersions is essential. However, such a distinction has been scarcely addressed both theoretically and experimentally to date. Notably, differences in the DOS and Landau level structures are expected to produce distinct experimental signatures. Further theoretical studies based on realistic models and first-principles calculations that investigate the tilting of Dirac cones in detail will be valuable for developing a more comprehensive understanding of these systems.




\appendix

%


\section{Density of states for characteristic Dirac systems}
\label{appenA}

In this appendix, we present expressions for the DOS for two Dirac-type systems that exhibit particle-hole symmetry. We use the 
energy, $\varepsilon$, with dimensions, instead of the dimensionless energy, $\tilde{\varepsilon}$, used in the main text.


%

\subsection{2D type-I Dirac fermions}

For 2D type-I Dirac fermions on a square lattice, where the first Brillouin zone is defined by $-\pi/a < k_x \leq \pi/a$ and $-\pi/a < k_y \leq \pi/a$, the DOS can be evaluated analytically for the case $\alpha = 0$ in Eq.~(\ref{model}). In this limit, the low-energy dispersion around each Dirac point is isotropic and linear, characterized by the Fermi velocity, $v_{\rm F}$. The resulting DOS reads
\begin{eqnarray}
D(\varepsilon) = \frac{N}{\pi} \left( \frac{a}{\hbar v_{\rm F}} \right)^2 |\varepsilon - \varepsilon_{\rm F}|. \label{d_0_1_1}
\end{eqnarray}
The overall factor is doubled due to the presence of two inequivalent Dirac points.

%

\subsection{2D narrow-sense type-III Dirac fermions}

\begin{figure}[bt]
\centering
\includegraphics[width=0.48\textwidth]{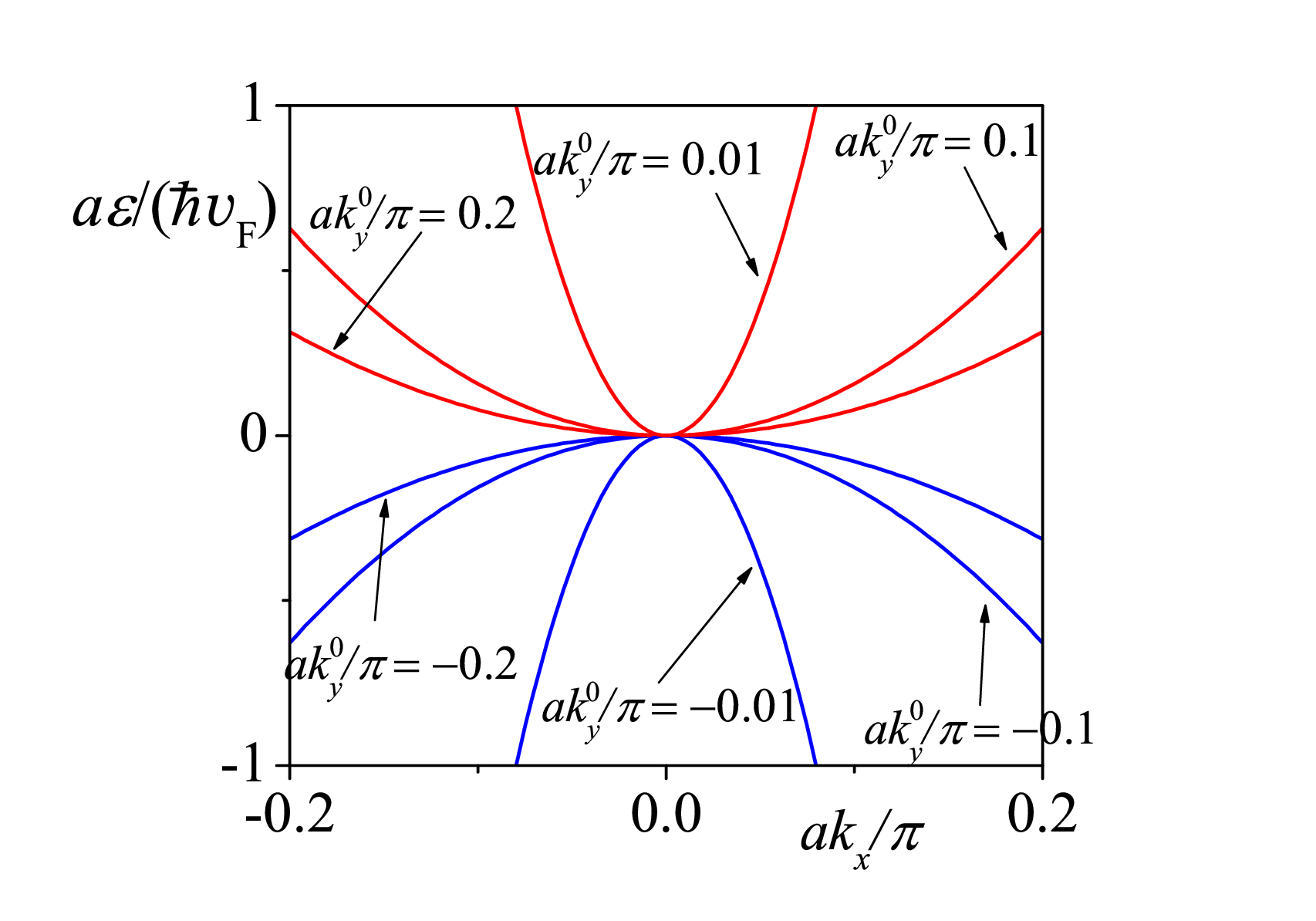}
\caption{(Color online)
Energy dispersion along the $k_x$-axis for several values of $k_y^0$, based on Eq.~(\ref{e1}), with $\varepsilon_{\rm F} = 0$. The red and blue curves correspond to the upper and lower bands, respectively.}
\label{fig_10}
\end{figure}

In the case of 2D narrow-sense type-III Dirac fermions, the dispersion is obtained by setting $\alpha = -\hbar v_{\rm F}$ in Eq.~(\ref{model}). Expanding around the Dirac point $(k_x, k_y) = (0, 0)$ yields
\begin{eqnarray}
\varepsilon \simeq \varepsilon_{\rm F} \pm \frac{\hbar v_{\rm F}}{2k_y^0} k_x^2, \label{e1}
\end{eqnarray}
at fixed $k_y = k_y^0$. This expansion is valid in the region $|k_x| \ll |k_y^0|$. When $k_y^0 = 0$, the dispersion becomes linear in $k_x$, i.e., $\varepsilon = \varepsilon_{\rm F} \pm \hbar v_{\rm F} |k_x|$. For $k_y^0 \neq 0$, the $k_x$-dependence resembles that of one-dimensional free electrons, with an effective mass controlled by $k_y^0$ [see Fig.~\ref{fig_10}].

By using Eq.~(\ref{model}), the constant-energy contour for the upper band is given by
\begin{eqnarray}
k_y = \frac{1}{2} \left( \frac{\hbar v_{\rm F}}{\varepsilon - \varepsilon_{\rm F}} k_x^2 - \frac{\varepsilon - \varepsilon_{\rm F}}{\hbar v_{\rm F}} \right). \label{FS}
\end{eqnarray}
The number of states [$\Omega$] between $\varepsilon_{\rm F}$ and $\varepsilon$ is obtained by integrating over the area enclosed by Eq.~(\ref{FS}) and the Brillouin zone boundary $k_y = \pi/a$. For $\varepsilon \simeq \varepsilon_{\rm F}$, this gives
\begin{eqnarray}
\Omega \simeq \frac{N}{3} \sqrt{ \frac{2a}{\pi \hbar v_{\rm F}} } \sqrt{ |\varepsilon - \varepsilon_{\rm F}| }.
\end{eqnarray}

Considering the existence of two Dirac points, the DOS in the narrow-sense type-III is 
\begin{eqnarray}
D(\varepsilon) &=& 2 \frac{d\Omega}{d\varepsilon} \nonumber \\
&=& \frac{N}{3} \sqrt{ \frac{2a}{\pi \hbar v_{\rm F}} } |\varepsilon - \varepsilon_{\rm F}|^{-1/2}, \label{d_0_3}
\end{eqnarray}
which diverges at $\varepsilon = \varepsilon_{\rm F}$.

%

%

\section{Analytical expression of electronic specific heat for generalized density of states}
\label{appenB}

We derive an analytical expression for the electronic specific heat, $C$, corresponding to Eq. (\ref{D1}) under the condition of 
half filling,  $\nu=1/2$. Starting from Eq.~(\ref{U_1}), the internal energy per site, $U$, can be rewritten in terms of $D(\varepsilon)$ as
\begin{eqnarray}
U= \frac{1}{N} \int_{-\infty}^{\infty} (\varepsilon - \mu) D(\varepsilon) f(\varepsilon) \, d\varepsilon + \mu \nu, \label{U_3}
\end{eqnarray}
where $\mu$ is the chemical potential.  Owing to the particle-hole symmetry of systems described by Eq.~(\ref{D1}), the chemical potential remains temperature-independent, i.e., $\mu = \varepsilon_{\rm F}$ for all $T$. The second term in Eq.~(\ref{U_3}) does not contribute to the specific heat. Consequently, it can be omitted in the following analysis.


Substituting Eq. (\ref{D1}) into Eq.~(\ref{U_3}) and changing variables to $x = (\varepsilon - \mu)/(k_{\rm B} T)$, we obtain
\begin{align}
C = k_{\rm B} D_0 \tilde{T}^{\beta+1} \int_{-\infty}^{\infty} \frac{|x|^{\beta+2} e^x}{(e^x + 1)^2} dx. \label{Ce_3}
\end{align}

The integral in Eq.~(\ref{Ce_3}) can be evaluated analytically as
\begin{align}
\int_{-\infty}^{\infty} \frac{|x|^{\beta+2} e^x}{(e^x + 1)^2} dx 
&= 2(\beta + 2) \sum_{k=0}^{\infty} \frac{(-1)^k}{(1 + k)^{\beta+2}} \int_0^\infty y^{\beta+1} e^{-y} dy \nonumber \\
&= 2(\beta + 2) \left(1 - 2^{-(\beta+1)}\right) \zeta(\beta + 2) \Gamma(\beta + 2), \label{Ce_4}
\end{align}
where $\zeta(x)$ and $\Gamma(x)$ are the Riemann zeta and gamma functions, respectively.

Substituting Eq.~(\ref{Ce_4}) into Eq.~(\ref{Ce_3}), we obtain the final expression for the electronic specific heat:
\begin{align}
C = 2 k_{\rm B} D_0 (\beta + 2) \left(1 - 2^{-(\beta + 1)}\right) \zeta(\beta + 2) \Gamma(\beta + 2) \left( \frac{\hbar v_{\rm F}}{a} \right)^{\beta+1} \tilde{T}^{\beta+1}. \label{CT}
\end{align}
This result confirms that the specific heat scales as $\tilde{T}^{\beta+1}$ for a DOS of the form $D(\varepsilon) \propto |\varepsilon - \varepsilon_{\rm F}|^\beta$.

\subsection{Electronic Specific heat for characteristic Dirac systems}

Using Eqs.~(\ref{d_0_1_1}) and (\ref{d_0_3}), we summarize the low temperature behavior of the electronic specific heat, $C$, for two Dirac-type systems:

\begin{itemize}
    \item \textbf{2D type-I Dirac fermions}:
    \begin{eqnarray}
    C &=& \frac{9}{2\pi} \zeta(3) \Gamma(3) k_{\rm B} \tilde{T}^2 \nonumber \\
    &\simeq& 3.444\, k_{\rm B} \tilde{T}^2, \label{C_D}
    \end{eqnarray}
    
    
    \item \textbf{2D narrow-sense type-III Dirac fermions}:
    \begin{eqnarray}
    C &=& \sqrt{\frac{2}{\pi}} \left(1 - 2^{-1/2} \right) \zeta(3/2) \Gamma(3/2) k_{\rm B} \tilde{T}^{1/2} \nonumber \\
    &\simeq& 0.5410\, k_{\rm B} \tilde{T}^{1/2}. \label{type3}
    \end{eqnarray}
    
    
\end{itemize}





\end{document}